\documentclass[reprint,prl,notitlepage,superscriptaddress, amsmath,amssymb]{revtex4-2}

\usepackage{float}
\usepackage{graphicx}
\usepackage{rotating}
\usepackage[toc,page]{appendix}
\usepackage{dcolumn}
\usepackage{bm}
\usepackage{soul}
\usepackage[colorlinks=true,allcolors=blue]{hyperref}
\usepackage{color}
\usepackage{xcolor}
\newcommand{\p}[1]{\textbf{#1}}
\newcommand{\Eqref}[1]{Eq.~\eqref{#1}}
\let\epsilon\varepsilon

\begin{document}

\title{Nonequilibrium nucleation theory for nonconserved fields:\\ from active matter to population dynamics}

\author{Michalis Chatzittofi}
\email{mc2623@cam.ac.uk}
\altaffiliation{These authors contributed equally to this work.}
\affiliation{DAMTP, Centre for Mathematical Sciences, University of Cambridge, Cambridge CB3 0WA, United Kingdom}
\author{Noah Ziethen}
\email{niz20@cam.ac.uk}
\altaffiliation{These authors contributed equally to this work.}
\affiliation{DAMTP, Centre for Mathematical Sciences, University of Cambridge, Cambridge CB3 0WA, United Kingdom}
\author{Cesare Nardini}
\affiliation{Service de Physique de l’\'Etat Condens\'e, CNRS UMR 3680, CEA-Saclay, 91191 Gif-sur-Yvette, France}
\affiliation{Sorbonne Universit\'e, CNRS, Laboratoire de Physique Th\'eorique de la Mati\'ere Condens\'ee, LPTMC, F-75005 Paris, France}
\author{Michael E. Cates}
\affiliation{DAMTP, Centre for Mathematical Sciences, University of Cambridge, Cambridge CB3 0WA, United Kingdom}

\date{\today}

\begin{abstract}
Classical nucleation theory (CNT) describes the formation of a stable phase from a metastable one. In equilibrium systems, it quantifies the free-energy competition between a favorable bulk gain and an unfavorable interfacial cost. 
For systems without detailed balance, the corresponding nonequilibrium nucleation theory (NNT) was so far developed only for cases with a conserved order parameter, such as active fluid-fluid phase separation. Here we construct the NNT for systems with a (single, scalar) nonconserved order parameter. Unlike in the conserved case, 
the nucleation barrier controlling (noise-driven) droplet growth is profoundly altered by
{\em deviations in the interfacial density profile} from the one arising during (deterministic) droplet relaxation. The barrier can nonetheless be analysed by carefully defining the reaction coordinate (droplet radius) to project out those deviations. We give explicit NNT predictions for models drawn from population dynamics and active matter, finding excellent agreement with numerical studies.
\end{abstract}

\maketitle
Rare events are crucial in settings ranging from solid state physics~\cite{brayTheoryPhaseorderingKinetics} and physical chemistry~\cite{allen2009forward} to finance~\cite{bouchaud1998langevin}, turbulence~\cite{falkovich1996instantons,ravelet2004multistability}, and geophysical flows~\cite{schmeits2001bimodal,ragone2018computation}. A prominent example is nucleation, where a stable phase is formed from a homogeneous metastable one, 
by noise-driven formation of a critical droplet which controls the least unlikely pathway between the two basins of attraction. Nucleation arises in systems as diverse as fluid-fluid demixing~\cite{brayTheoryPhaseorderingKinetics,onuki2002phase}, ferromagnetism~\cite{ishibashi1971note,rikvold1994metastable}, crystallization~\cite{vekilov2010nucleation}, phase-ordering systems~\cite{brayTheoryPhaseorderingKinetics,onuki2002phase}, reaction-diffusion systems~\cite{hinrichsen2000non,elgartRareEventStatistics2004, Halatek2018May,hellerPatternStabilityReactiondiffusion2026,cardy1998field}, and ecology~\cite{bastiaansenMultistabilityModelReal2018,Korniss2005Mar,DeAngelis2001Jan,Gandhi1999Sep,tanakaSpatialGeneDrives2017,giometto2021antagonism}. 
Nucleation and structure formation in such cases, especially far from equilibrium, have been under increasing investigation
using theory~\cite{tanakaSpatialGeneDrives2017,meerson2011extinction,schuttlerEffectsPhaseSeparation2024,escudero2004extinction,bastiaansenMultistabilityModelReal2018,agranov2021extinctions,maUniversalityNoiseinducedResilience2021,lavrentovich2019nucleation,andreghetti2025enzyme}, natural observations~\cite{bastiaansenMultistabilityModelReal2018,jablonski2008extinction}, and experiments~\cite{giometto2021antagonism}. In the realm of active matter, nonequilibrium nucleation has been argued to make the active nematic transition re-entrant at low noise~\cite{shankar2018defect}; to destroy the ordered phase of polar flocks~\cite{benvegnen2023metastability}; to sustain bubbly phase-separation~\cite{Cates2025May,catesClassicalNucleationTheory2023}, and to influence the sizes of biomolecular condensates~\cite{lee2023size}. 

In equilibrium systems described by a single order parameter, and in the absence of spatial anisotropies, the nucleation of the stable phase from a metastable one generally proceeds by formation of a spherical droplet which, once it has reached the critical radius $R_\mathrm{c}$, relaxes deterministically to form the stable phase. (For smaller sizes, only the noise prevents it relaxing instead to $R=0$.)
This pathway is described by Classical Nucleation Theory (CNT), in which the probability (or rate) of nucleation is~\cite{oxtobyHomogeneousNucleationTheorya, DebenedettiPabloG2020, karthikaReviewClassicalNonclassical2016b} 
$\mathbb{P} \asymp \exp\left(-U_\mathrm{eq}(R_\mathrm{c})/k_\mathrm{B}T\right)$. Here, $\asymp $ stands for logarithmic equivalence~\cite{Touchette2009Jul} in the small temperature limit, and $k_\mathrm{B}$ is Boltzmann's constant. In three spatial dimensions, the free energy barrier is found as
\begin{align}\label{eq:DeltaF-eq}
U_\mathrm{eq}(R_\mathrm{c})
= \frac{4\pi}{3}\sigma_\mathrm{eq} R_{\mathrm{c}}^2 + \mathcal{O}(R_\mathrm{c},T)\qquad d=3\,,
\end{align}
where $\sigma_\mathrm{eq}$ is the interfacial tension between phases. The critical radius $R_\mathrm{c}$ is determined by the supersaturation in systems with a conserved order parameter, such as in phase-separating fluids, and by the free energy difference between the stable and the metastable state when the order parameter is not conserved~\cite{brayTheoryPhaseorderingKinetics}. 

Investigating rare events with equilibrium dynamics is significantly easier than when detailed balance is broken, for two reasons. First, $U_{\mathrm{eq}}(R_\mathrm{c})$ can be computed by comparing the free-energy of the initial metastable state with that of the lowest saddle that allows the system to escape its basin of attraction ({\em i.e.}, the droplet of radius $R_\mathrm{c}$). Second, detailed balance requires that the least unlikely trajectory from the metastable state to the saddle, called the instanton, is the time-reversal of the relaxation path from the saddle to the metastable state~\cite{Touchette2009Jul,Freidlin1998,Bouchet2016Jun}. Neither of these simplifications apply once detailed balance is broken: free-energy arguments cannot be used to estimate the probability of rare events, and one does not know {\em a priori} the instanton trajectory. 
Consequently, computing rare event rates in non-equilibrium systems is impossible analytically (with very few exceptions~\cite{bodineau2005current,mallick2022exact}). Instead,
sophisticated rare-events algorithms have been developed to do so numerically~\cite{bucklew2004introduction,giardina2006direct,lecomte2007numerical,heymann2008pathways,vanden2012rare,grafke2015instanton,grafke2019numerical,nemoto2014computation,ferre2018adaptive,yan2022learning,PhysRevResearch.6.043110,PhysRevX.13.041044}. 

\begin{figure}
    \centering    \includegraphics[width=\linewidth]{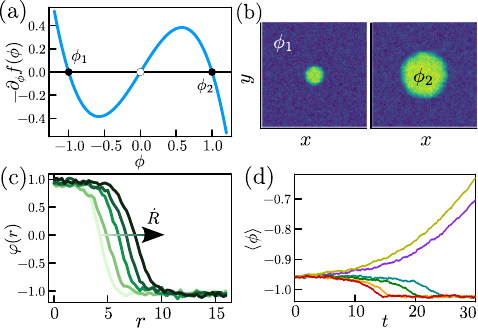}
    \caption{(a) Plot of $\langle\dot\phi\rangle =-\mu$ against $\phi$ for spatially homogeneous states of Active Model A (AMA). The two states $\phi_{1,2}$ are locally stable and are denoted by dark circles. (b) Time evolution of a nucleating droplet of stable phase $\phi_2$ in the metastable phase $\phi_1$. (c) The droplet expands with a velocity $\dot{R}$. The plot reports the time-evolution of the interfacial profile (colors from light to dark) as obtained from simulations of AMA. (d) Trajectories of the global density during two successful and four unsuccessful nucleation events; different lines correspond to different realizations of the stochastic process starting from a spherical droplet with $R\lesssim R_\mathrm{c}$. Data for panels (b--d) corresponds to simulations of AMA in $d=2$, with $D=1$, $K=1$, $f(\phi)=-\phi^2/2+\phi^4/4$, $\lambda=-0.5$, $h=0.1$.}
    \label{fig:fig1}
\end{figure}

Despite these issues, CNT has been used phenomenologically, with some numerical evidence of success, to address active phase-separating systems, both at fixed particle number~\cite{richard2016nucleation,redner2016classical,levis2017active} and in subcellular liquid-liquid phase separation for chemically reacting species~\cite{cho2023tuning,shimobayashiNucleationLandscapeBiomolecular2021b}. Furthermore, CNT was used for a minimal model of biomolecular condensates in a specific regime that allows a mapping to an effective equilibrium system~\cite{Ziethen2024Jun,noahprl}. So far, however, the well-founded extension of the CNT formalism to create a nonequilibrium nucleation theory (NNT) has been limited to the case of conserved dynamics --- specifically to Active Model B+, which minimally describes phase-separation in nonequilibrium systems with a (single, scalar) conserved order parameter~\cite{catesClassicalNucleationTheory2023}. 
Thus it remains unclear whether NNT can be extended to the case of nonconservative dynamics as governs many reaction-diffusion type systems ranging from biomolecular condensates to ecology. 

In this Letter, we develop and apply NNT for field theories with a (single, scalar) nonconserved order parameter. We do so by systematically expanding the stochastic dynamics of the droplet radius $R(t)$ in the form
\begin{align}\label{eq:Rdot}
\dot R = \mathcal{M}(R) \left[-\partial_{R} U(R) \right]  + \sqrt{2 T\mathcal{M}(R)} \zeta,
\end{align}
where $\zeta$ is a Gaussian noise with zero mean and variance $\langle \zeta(t)\zeta(t')\rangle = \delta(t-t')$, $U(R)$ is the quasipotential, $\mathcal{M}(R)$ is the (emergent) mobility of a reaction coordinate $R$, and $T$ quantifies the noise strength, assumed small in the CNT/NNT limit. This allows us to analytically obtain the critical radius $R_\mathrm{c}$ and the nucleation rate set by the energy barrier $U(R_\mathrm{c})$. Because in Eq.~\eqref{eq:Rdot} the droplet radius $R$ is a 1D reaction coordinate, its dynamics resembles an equilibrium system with detailed balance. This is however deceptive: we will show that the interfacial profile of the noise-driven growing droplet {\em cannot} be equated to the one that prevails during the relaxation dynamics. 
This is because the mobility $\mathcal{M}(R)$ has to be calculated with particular care. 

After establishing NNT in the form \eqref{eq:Rdot}, we apply it to two models from population dynamics~\cite{stephens1999allee} and to Active Model A (AMA). The latter describes active non-conserved relaxation between competing homogeneous states~\cite{PhysRevLett.124.240604}; and can be viewed as a field theory for kinetic Ising models that violate detailed balance at lowest (deterministic) order.
For AMA we validate our NNT predictions by minimizing numerically the Freidlin--Wentzell action of the original field theory. Further details are presented in a companion paper~\cite{pre}.

The class of models investigated in this Letter involve a single scalar order parameter $\phi(\p{x},t)$ obeying a local and isotropic dynamics in $d\ge 2$ dimensions:
\begin{align}\label{eq:model}
    \partial_t \phi &= -\mu[\phi] + \sqrt{2T D(\phi)}\xi\,,
\end{align}
where $\xi$ is a zero-mean white Gaussian noise with  $\langle \xi(\p{x},t) \xi(\p{x}',t') \rangle = \delta(\p{x}-\p{x}') \delta(t-t')$. We choose $\mu[\phi]$ such that without noise \eqref{eq:model} admits two uniform phases as stable fixed points, $\phi(\p{x}) = \phi_{1,2}$, with $\phi_1$ the metastable phase and $\phi_2$ globally stable. 
We study the nucleated transition from the metastable to the stable phase. Fig.~\ref{fig:fig1} illustrates this for AMA, which we return to below. 

Although our results can be extended to general functionals $\mu[\phi]$, we will assume here that $\mu$ is a local function of $\phi$ and its derivatives, while $D(\phi)>0$ is a local function of $\phi$ only. 
Equilibrium Model A (in the Hohenberg-Halperin classification~\cite{hohenbergTheoryDynamicCritical1977,onuki2002phase}) corresponds to $\mu=D\delta\mathcal{F}/\delta\phi$ with $\mathcal{F}[\phi]= \int (f(\phi)+K |\nabla\phi|^2/2)\,d{\bf x}$ and $f(\phi)$ a double-well local free energy. 
Importantly, we also include non-equilibrium models where $\mu/D$ is not the functional derivative of any $\mathcal{F}[\phi]$. (An example is Active Model A, where $\mu/D = \delta\mathcal{F}/\delta\phi+\lambda|\nabla\phi|^2$.)  

We address the weak noise limit where droplets of the stable phase, when smaller than $R_\mathrm{c}$ shrink deterministically with a velocity $v_0(R,R_\mathrm{c})$, which we assume to be small. (Equivalently, $R$ and $R_\mathrm{c}$ are large.)
These are standard CNT assumptions which, in equilibrium models,  require the metastable and stable uniform states to have almost equal free energy densities. We restrict to cases in which fluctuations away from a spherical droplet shape are resisted by a positive capillary-wave interfacial tension $\sigma_\mathrm{cw}$~\cite{pre}. Such fluctuations are then of order $\delta R(\Omega,t)\sim \sqrt{T/\sigma_\mathrm{cw}}\ll R$, with $\Omega$ the angular coordinates, and negligible in the CNT/NNT limit. 

A stronger assumption, justified below, is that
the droplet density profile for configurations on the instanton trajectory ($\dot R> 0$) remains close to the interfacial profile $\varphi(r,R(t))$ for a noiseless droplet of the same radius undergoing relaxation dynamics ($\dot R<0$). Here the droplet is centred at the origin and $r=|{\bf x}|$. Thus we assume  
\begin{align}
\epsilon({\bf x},R,t) := \phi({\bf x},t)-\varphi(r,R(t))\label{epsdef}
\end{align}
is small. But, unlike in CNT, we {\em do not} suppose $\epsilon = 0$ and indeed will show that doing so gives wrong results.

We use below the notation 
$\varphi(r,R) = \varphi_0(r-R)+(1/R)\varphi_1(r-R)+\mathcal{O}(1/R^2)$, in which $\varphi_0$ is the density profile of a flat interface, and $\varphi_1$ is a correction due to curvature. 
Without loss of generality, for any spherically symmetric $\phi(r)$ we will also below write $\mu[\phi] = \mu_0[\phi]+r^{-1}\mu_1[\phi] + ...$ where $\mu_0$ and $\mu_1$ depend on $r$ only via $\phi$, and the omitted terms vary as $1/r^n$ with $n\geq 2$.

To obtain the correct dynamics of $R(t)$, a crucial step is to carefully define the droplet radius $R[\phi]$ for a given field configuration $\phi({\bf x})$. There are many possible definitions that differ by $\mathcal{O}(1)$ in the large $R$ limit and, at first glance, any of these would do. But we shall make a special choice for which the projection from the stochastic dynamics of $\phi$ in~\eqref{eq:model}, to that of $R$ in~\eqref{eq:Rdot}, avoids finding $\epsilon$ in~\eqref{epsdef}.  

Thus, echoing earlier works on moving interfaces in reaction-diffusion systems~\cite{kuramoto1980instability,kawasaki1982kinetic} and capillary waves in equilibrium models~\cite{kawasaki1982kinetic-i,bausch1991effects}, we define $R(t)$ as the (unique~\footnote{It is easy to show that the derivative of Eq. \eqref{eq:defn} with respect to $R$ is positive as far as $\epsilon$ is small and $\psi'\propto\varphi'$, which covers all cases considered. This proves that, at least in the small-$\epsilon$ regime, Eq. \eqref{eq:defn} is satisfied for a single value of $R$.}) solution of 
\begin{align}\label{eq:defn}
    0 = \int_{\bf x}\; \epsilon({\bf x},R,t) \psi'(r-R).
\end{align}
Here $\int_{\bf x}$ stands for the integral over $\bf x$ and $\psi'(u)$ is an auxiliary function peaked at $u=0$. So far, $\psi'$ is arbitrary but it is chosen below to eliminate $\epsilon$ as promised above. 
Taking the time derivative of~\eqref{eq:defn} we find
\begin{align}\label{eq:Rdot-exact}
     \dot R(t) = -\frac{\int_{\bf x}\;\psi'\dot\phi}{\int_{\bf x}  \psi'\varphi' - \int_{\bf x} \psi'' \epsilon}.
\end{align}
Substituting in Eq.~\eqref{eq:model}, then expanding for large $R$, small $\epsilon$, and small velocity $v_0$ of the flat interface, and finally performing the angular integral, we find
\begin{align}\label{eq:Rexp}
    \dot R&=
    \frac{\int_r \, \psi'( \mu_1[\varphi_0]+\mathcal{L}\varphi_1)}{\int_r \psi' \varphi_0'}
    \left[
    \frac{\int_r \, \psi' \mu_0[\varphi_0]}{\int_r \,  \psi'(\mu_1[\varphi_0]+\mathcal{L}\varphi_1)}
    +\frac{1}{R}
    \right] \nonumber \\
    & +\sqrt{2 T\mathcal{M}(R)} \zeta(t)+\mathcal{O}(...) + C[\phi]\,.
\end{align}
Here $\mathcal{L}(\cdot)$ is the operator $\mu_0$ linearized around $\varphi_0$, and the mobility, which at this stage depends on $\psi'(u)$, obeys 
\begin{align}
    \mathcal{M}(R) &= \frac{\int_r \, \psi'^2 D(\varphi_0)}{R^{d-1} S_d\left[\int_r \, \psi' \varphi_0'\right]^2} \,.\label{eq:M-generic}
\end{align} 
The higher order terms in \eqref{eq:Rexp} are of order $\mathcal{O}(...)=\mathcal{O}(R^{-2},v_0 R^{-1}, \sqrt{T}R^{-(d+1)/2}, \epsilon^2, v_0 \epsilon, \sqrt{T}\epsilon)$ and are irrelevant in the CNT/NNT limit under study. Crucially however, \eqref{eq:Rexp} is not yet a closed equation for the droplet radius, because of the final term 
$C[\phi] := \left(\int_\p{x}\,\, \psi' \mathcal{L} \epsilon\right) / \int_\p{x} \, \psi'\varphi_0'$. This depends not only on the field $\phi({\bf x)}$ via $\epsilon$, but on the choice of auxiliary function $\psi'$ and hence on the precise definition of $R[\phi]$. Yet two definitions of $R$ that differ by $\mathcal{O}(1)$ in the large $R$ limit must, physically, share the same Langevin equation \eqref{eq:Rdot}.

This means that for large droplets there must be a conspiracy between the $C$ term and the mobility $\mathcal{M}$ (which also is $\psi'$-dependent) to arrive finally at a definition-independent Langevin equation for $R(t)$.
Fortunately however, explicit calculation can be avoided, because $C=0$ exactly when $\psi' \in \mathrm{Ker}(\mathcal{L}^\dagger)$. Making this choice
we obtain Eq.~\eqref{eq:Rdot} where $\mathcal{M}(R)$ obeys Eq.~\eqref{eq:M-generic}, the critical radius is
\begin{align}\label{eq:Rc-generic}
R_\mathrm{c} &=\frac{(d-1)\sigma}{
\int_r \, \psi' \mu_0[\varphi_0]}\,,
\end{align}
and the quasipotential is 
\begin{align}\label{eq:U+}
    U(R) =    \frac{S_d R^{d-1}\sigma  
    \int_r \, \psi' \varphi_0' }{\int_r \, \psi'^2 D(\varphi_0)}\left( 1 - \frac{R (d-1)  }{R_\mathrm{c} d } \right)\,.
\end{align}
Above we have introduced the NNT interfacial tension
$\sigma = -\int_r \,  \psi'\mu_1[\varphi_0]/(d-1)$.
Note that the drift velocity of a noiseless flat interface obeys $v_0 = \int_r \, \psi' \mu_0[\varphi_0] / \int_r \psi'\varphi_0'$ so that, as expected, small $v_0$ corresponds to large $R_\mathrm{c}$. (Both $v_0$ and $R_\mathrm{c}$ must be positive for nucleation to occur.) 

We show finally that $\epsilon$ is small as assumed above. Expanding \Eqref{eq:model} in $\epsilon$ we get
\begin{align}\label{eq:pteps}
(\partial_t + \mathcal{L}_\varphi )\epsilon = -\mu[\varphi] +\varphi' \dot{R}+\sqrt{2TD(\varphi)}\xi + \mathcal{O}(\epsilon^2,\epsilon\sqrt{T}),
\end{align}
where $\mathcal{L}_\varphi$ is the operator $\mu$ linearized around $\varphi$. Notice that we are interested only in estimating the order of magnitude of $\epsilon$ at $r\sim R$ because it is the only quantity that enters in Eq.~\eqref{eq:Rdot-exact} after expanding in $\epsilon$.
Neglecting for now $\partial_t\epsilon$ on the left hand side, and observing that $\mu[\varphi]\sim \mathcal{O}(v_0, 1/R)$, we find that $\epsilon\sim \mathcal{O}(v_0,\dot{R},1/R,\sqrt{T})$. Then, using that $\dot R\sim \mathcal{O}(v_0,1/R)$, we have that $\epsilon\sim\mathcal{O}(v_0,1/R,\sqrt{T})$.
We can further show from \eqref{eq:pteps} that $\partial_t\epsilon\sim\mathcal{O}(v_0\dot R, {\dot R}^2, \ddot{R}, \sqrt{T} \dot{R})$ which is negligible for estimating $\epsilon$. This concludes the argument: all the $\mathcal{O}(...)$ terms in \Eqref{eq:Rexp} are indeed negligible as we proposed. For further details, see~\cite{pre}. 

The derivation of Eqs.~(\ref{eq:Rdot}, \ref{eq:M-generic}-\ref{eq:U+}) represents the main analytical result of this Letter. Surprisingly, even when activity only enters via deterministic driving terms in \eqref{eq:model} via $\mu$, it contributes to the fluctuation dynamics of $R(t)$ through the dependence of the mobility in \eqref{eq:M-generic} on $\psi'$. It also follows from \Eqref{eq:U+} that the quasipotential barrier $U(R_\mathrm{c})$ cannot be expressed solely as a function of $R_\mathrm{c}$ and the tension $\sigma$, as was true in equilibrium CNT~\cite{onuki2002phase,brayTheoryPhaseorderingKinetics} and in conserved active NNT~\cite{catesClassicalNucleationTheory2023,Cates2025May}.

Furthermore, although we obtained the Langevin equation for $R(t)$ without computing $\epsilon$, we now show that the distortions  it encodes (denoted hereafter $\epsilon_\mathrm{A}$) of the interfacial profile on the instanton trajectory, relative to that on the relaxation path, crucially alter the energy barrier. This contrasts with both equilibrium CNT and the conserved NNT, where $\epsilon_\mathrm{A}$ does not affect (to the order considered) the nucleation barrier~\cite{catesClassicalNucleationTheory2023,pre}.
\begin{figure}[t]
    \centering
\includegraphics[width=\columnwidth]{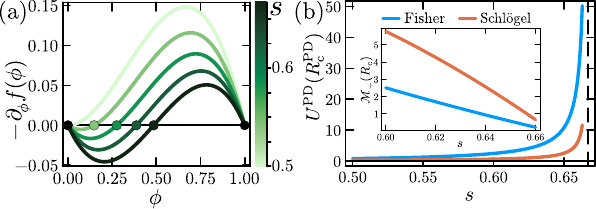}
    \caption{(a) The driving force for the two population dynamics models, showing the two fixed points of the deterministic dynamics $\phi_1=0$, $\phi_2=1$; the asymmetry between the two species $s$ modifies the location of the unstable fixed point.  
    (b) Barrier heights for different values of asymmetry $s$ showing how nucleation probabilities are affected by the form of the noise variance; the inset reports the mobilities $\mathcal{M}$ for the two models.}
    \label{fig:fig3}
\end{figure}

To show this, we restrict to constant $D$ (setting $D=1$), and
start from the Freidlin-Wentzell (FW) action~\cite{Freidlin1998} associated with Eq.~\eqref{eq:model}, $\mathcal{A} = \frac{1}{4T}\int dt\int_{\bf x}  (\dot \phi +  \mu)^2$. Evaluating $\mathcal{A}$ for $\phi({\bf x},t)=\varphi(r,R(t))$ and expanding for large $R$ and small $v_0$, we obtain \Eqref{eq:Rdot} with $\mathcal M$ and $U$ in (\ref{eq:M-generic},\ref{eq:U+}) respectively replaced by~\cite{pre}
\begin{align}
    \mathcal{M}_\varphi(R) &= \frac{1}{R^{d-1}S_d}  \frac{1}{\int_r \,\varphi_0'^2}\label{eq:Munp}\\
    U_\varphi(R) &=    \frac{-S_d R^{d-1} \int_r \varphi_0'(\mathcal{L}\varphi_1+ \mu_1 )}{(d-1)}\left( 1 - \frac{R (d-1)  }{R_\mathrm{c} d } \right).\label{eq:Uphi}
\end{align}
These results assume that the instanton follows a time-reversed relaxation (TRR) trajectory; they can also be found by setting $\psi'=\varphi_0'$ in (\ref{eq:M-generic},\ref{eq:U+}). 
In equilibrium systems, \eqref{eq:Munp} and \eqref{eq:Uphi} are correct, because time-reversal symmetry of \eqref{eq:model} ensures $\epsilon_A=0$~\cite{Touchette2009Jul,Freidlin1998,Bouchet2016Jun} and the TRR ansatz holds. Moreover, translational invariance and the self-adjointness of $\mathcal{L}$ then imply $\mathcal{L}^\dagger \varphi_0'=0 $. Thus $\varphi_0' \in \mathrm{Ker}(\mathcal{L}^\dagger)$ and
the choice $\psi'=\varphi_0'$ is also correct.

\begin{figure}[t]
    \centering
\includegraphics[width=1\linewidth]{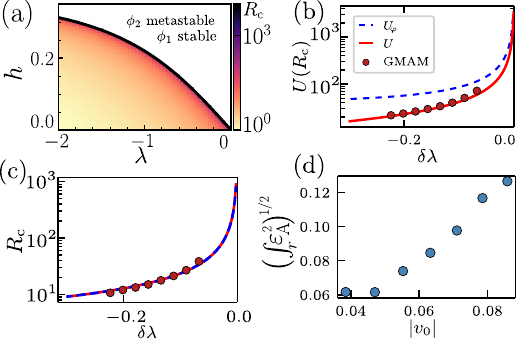}
    \caption{Results for AMA in two dimensions: (a)  Color map of the critical radius $R_\mathrm{c}$ in the NNT limit as a function of activity $\lambda$ and external field $h$: $\phi_1$ ($\phi_2$) is metastable for finite (infinite) values of $R_\mathrm{c}$, and NNT is valid close to the solid line (where $R_\mathrm{c}\to\infty)$. (b) The quasipotential barrier in the NNT analysis (orange line); in the time-reversed relaxation (TRR) ansatz $\epsilon_A\to 0$ (blue dashed line): and by numerical (GMAM) minimization of the action (dots). (c) The same comparison for the critical radius. (d) Deviation in the interfacial profile between NNT and TRR, quantified via $({\int_r \epsilon_\mathrm{A}}^2)^{1/2}$, showing consistency with  $({\int_r \epsilon_\mathrm{A}^2})^{1/2}\sim v_0$ as NNT predicts. In (b--d) data is shown at fixed $h=0.2$ and varying $\lambda =\lambda_*(h)+\delta\lambda$ where $\lambda_*(0.2)=-0.8768$. 
    See \cite{pre} for further details.
    }
    \label{fig:fig2}
\end{figure}

For generic non-equilibrium systems, however, that choice is wrong: $\mathcal{M}$ and $U$ differ from $\mathcal{M}_\varphi$ and $U_\varphi$~\footnote{The expression we obtain for $R_\mathrm{c}$ here is $R_\mathrm{c} =\frac{(\int_r \varphi_0'\mathcal{L}\varphi_1+ \mu_1 \varphi'_0)}{\int_r \mu_0\varphi_0'}$ which looks different from \Eqref{eq:Rc-generic}. Yet, these two expressions are equivalent (up to terms of order $\mathcal{O}(R_\mathrm{c}^0)$) as follows from the fact that $R_\mathrm{c}$ depends on the noiseless dynamics only. See~\cite{pre} for details.}. Moreover the difference in $U$ is not $\mathcal{O}(\epsilon)$ but $\mathcal{O}(1)$, for example $U_\varphi(R_\mathrm{c})$ overestimates the barrier height $U(R_\mathrm{c})$ by more than a factor 2 in AMA (see Fig.~\ref{fig:fig2}). We show in~\cite{pre} how the minimization of the action can be achieved without the TRR ansatz, thereby confirming Eqs.~(\ref{eq:M-generic}--\ref{eq:U+}). 

We now apply our NNT results to some specific models. 
First we consider reaction-diffusion models~\cite{tauber2014critical,hinrichsen2000non} of a type used in population dynamics~\cite{garcia2012noise,korolevGeneticDemixingEvolution2010} to describe strong Allee effects~\cite{stephens1999allee}. The order parameter $\phi\in [0,1]$ is a normalized population density, obeying~\eqref{eq:model} with 
\begin{align}\label{eq:model_react_diff}
    \mu[\phi] = \partial_\phi f_\mathrm{PD} - K \nabla^2 \phi 
\end{align}
where $f_\mathrm{PD}(\phi)$ is a local function of $\phi$, and $K$ is a spatial diffusivity. 
We will specifically consider a Fisher-type equation~\cite{murray2003mathematical} that was recently used to describe gene-drive dynamics~\cite{tanakaSpatialGeneDrives2017}, which has $\partial_\phi f_\mathrm{PD}(\phi) = s\phi(\phi-1)(\phi - \phi_s)$, with $\phi_s = (2s-1)/s$; here, $\phi$ is the fraction of gene-drive alleles, and $s$ is the selective advantage. 
The transition between uniform states $\phi_1=0$ (metastable) and $\phi_2 = 1$ (stable) is by nucleation for $1/2 < s < 2/3$.

Whereas $\mu=\delta\mathcal{F}/\delta 
\phi$ with $\mathcal{F}= \int_{\bf x}(f_\mathrm{PD} + K|\nabla\phi|^2/2)$, multiplicative noise breaks detailed balance and places these models far from equilibrium. We consider two noise types: a Schl\"{o}gl form having $D(\phi) = s\phi(\phi+1)(\phi_s + \phi) + D_0$~\cite{schloglChemicalReactionModels1972,schuttlerEffectsPhaseSeparation2024,ewensMathematicalPopulationGenetics2004}, and a 
Fisher-type noise $D(\phi) = \phi(1-\phi) + D_0$~\cite{korolevGeneticDemixingEvolution2010,tanakaSpatialGeneDrives2017}. In both cases, the constant term $D_0\ll 1$ prevents absorption at $\phi=\phi_{1,2}$. 
In Fig.~\ref{fig:fig3}(b) we plot nucleation barriers for both cases. 
The quasipotential $U$ depends on the form of the noise $D(\phi)$; therefore, unlike in equilibrium, it cannot be inferred (up to the scale factor $T$) from the noiseless dynamics alone
~\cite{tanakaSpatialGeneDrives2017}. 
(See~\cite{pre} for details, and numerical verification.)

Our final example is Active Model A (AMA)~\cite{PhysRevLett.124.240604}: this is one of the simplest field theories for nonequilibrium phase ordering.
The AMA equation of motion takes the form \eqref{eq:model} with $D=1$ and 
\begin{align}\label{eq:ama}
    \mu[\phi] = \partial_\phi f(\phi) -  
\nabla^2 \phi + \lambda (\nabla\phi)^2
\end{align}
where $f(\phi)=-\phi^2/2 + \phi^4/4+h \phi$.
The term in $\lambda$ breaks detailed balance at lowest-order in the Landau Ginzburg expansion; it cannot be written $\delta \mathcal{F}/\delta\phi$ for any $\mathcal{F}$. The parameter $h$ can be viewed as an external field, and the limit $\lambda=0$ is the long-studied equilibrium Model A~\cite{hohenbergTheoryDynamicCritical1977}, whose stationary states are $\phi_{1,2} = \pm 1 - h/2 + \mathcal{O}(h^2)$ (as shown for $h\to 0$ in Fig.~\ref{fig:fig1}(a)). For positive $h$ (say), the phase with $\phi_1<0$ is then metastable and $\phi_2$ is stable. 

Our results can be applied to AMA by noticing~\cite{pre} that $\mu_1[\varphi_0]=-(d-1)\varphi_0'$ and choosing $ \psi' = e^{-2\lambda \varphi_0} \varphi_0'$. Then, as required, $ \psi'\in \mathrm{Ker}(\mathcal{L}^\dagger)$. The function $\psi(\phi)$ is known as the pseudo-density, and was previously introduced in the context of conserved active field theories~\cite{solon2018generalized,Cates2025May,PhysRevX.8.031080}. Although also deployed in NNT for AMB+~\cite{catesClassicalNucleationTheory2023}, its use was avoidable there since the TRR ansatz gives identical results to the required order (see~\cite{pre} for an explanation).
As illustrated in Fig.~\ref{fig:fig2}, TRR does not hold for AMA.

In AMA, in contrast with the equilibrium case ($\lambda=0$) and the population dynamics models considered above, the relative stability of $\phi_{1,2}$ involves the activity $\lambda$ as well as the local free energy $f(\phi)$. When $h>0$ and $\lambda=0$, $\phi_1$ is the stable state and $\phi_2$ is metastable, but at fixed $h$ and sufficiently negative $\lambda$, the reverse is true (see Fig.~\ref{fig:fig2}).

The critical radius $R_\mathrm{c}$ diverges at the switchover point $\lambda_*(h)$, which is the solid line in Fig.~\ref{fig:fig2}(a). 
Just as with CNT, our NNT holds formally for large $R_\mathrm{c}$, that is, close to this line. We report in Fig.~\ref{fig:fig2}(b) the effective energy barrier for nucleation $U(R_\mathrm{c})$ and also the barrier $U_\phi(R_\mathrm{c})$ found via the TRR ansatz. This ansatz as in \eqref{eq:Uphi} overestimates $U(R_\mathrm{c})$ by an $\mathcal{O}(1)$ factor, giving nucleation times that are infinitely too long for $T\to 0$. This discrepancy disappears only close to equilibrium
where both barriers are approximated to leading order in $(\lambda,h)$ by $U(R_\mathrm{c}) \simeq U_\varphi(R_\mathrm{c}) =-20\,\pi(12\lambda +45h)^{-1}$~\cite{pre}. In Fig.~\ref{fig:fig2}(c)
we show that, in contrast, our NNT and the TRR ansatz give the same critical radius $R_\mathrm{c}$. Indeed, unlike the barrier height, $R_\mathrm{c}$ is noise-independent: it locates the unstable fixed point of the noiseless dynamics. 

The validity of our theory is confirmed by minimizing the Freidlin--Wentzell action for AMA numerically. This is done by combining a geometric minimum action method (GMAM)~\cite{heymann2008,
Grafke2019Jun} with a Ritz method~\cite{PhysRevResearch.2.033208} using finite-difference
discretization in space and a Chebyshev basis in time. Details are given in~\cite{pre}, as are similar results for a population model. Fig.~\ref{fig:fig2}(b,c) show excellent agreement with our analytical predictions for the critical radius and the nucleation barrier, whereas $U_\varphi$ clearly overestimates the barrier. 
We report in Fig.~\ref{fig:fig2}(d) numerical values of
the mean-square difference in the interfacial profile between our NNT and the TRR ansatz. As we predicted analytically for $\lambda\neq 0$, this scales as $\epsilon_\mathrm{A}\sim \mathcal{O}(v_0)$.

Concluding, we have systematically extended Classical Nucleation Theory to address non-equilibrium field theories with a (single, scalar) {\em non-conserved} order parameter. The resulting Nonequilibrium Nucleation Theory is considerably more subtle than that for the conserved case~\cite{catesClassicalNucleationTheory2023} because deviations in the interfacial profile from that on the time-reversed relaxation path now generically enter at leading order when calculating the nucleation barrier $U(R_\mathrm{c})$. Our results were made possible by carefully defining the reaction coordinate (droplet radius) so that these deviations get projected out of the Langevin equation for the radial coordinate. The latter then unambiguously yields the quasipotential at CNT/NNT level.
For Active Model A, we confirmed our results numerically, finding excellent agreement. 

We have exemplified our general theory with cases drawn from active matter and population dynamics. Yet, several other applications are natural. We expect for example the approach developed here to be applicable to nucleation of biomolecular condensates~\cite{shimobayashiNucleationLandscapeBiomolecular2021b,lee2023size,wilken2024nucleation} beyond the specific parameter regimes that allow a mapping to an effective equilibrium model~\cite{noahprl,Ziethen2024Jun}. Likewise we expect that  --- beyond the minimal setups considered in this Letter --- our theory will be extendable to describe nucleation in ecological contexts~\cite{tanakaSpatialGeneDrives2017,schuttlerEffectsPhaseSeparation2024,escudero2004extinction,bastiaansenMultistabilityModelReal2018,maUniversalityNoiseinducedResilience2021,lavrentovich2019nucleation,bastiaansenMultistabilityModelReal2018,jablonski2008extinction,giometto2021antagonism}, in reaction-diffusion systems~\cite{meerson2011extinction,agranov2021extinctions,andreghetti2025enzyme}, chemical kinetics, and active systems with multiple order parameters~\cite{PhysRevE.111.034124,PhysRevLett.134.117103,shankar2018defect,benvegnen2023metastability}.
Our results may also be used to benchmark and complement machine learning or optimization techniques for computing rare events~\cite{Grafke2019Jun,PhysRevX.13.041044,PhysRevResearch.2.033208,Simonnet2023Oct,PhysRevLett.133.038301}.

\begin{acknowledgements}
We thank Ronojoy Adhikari, Filippo De Luca, Robert Jack, and Nicolas Valade for fruitful discussions. We thank J. Tailleur for bringing~\cite{kuramoto1980instability,kawasaki1982kinetic,kawasaki1982kinetic-i,bausch1991effects} to our attention while this work was being finalized. CN acknowledges the
support of the ANR grant PSAM, and the support of the INP-IRP grant IFAM. This research was supported in part by grant NSF PHY-2309135 to the Kavli Institute for Theoretical Physics (KITP).
We acknowledge the support and funding in part by the EPSRC through grant EP/Z534766/1.

\end{acknowledgements}
\bibliography{bibtex}

\end{document}